\documentclass{ceab}   %% loading ceab.cls creates CEAB style

\usepackage{epsfig}     %\  Used to include figures
\usepackage{graphicx}   %/

\usepackage{ceabbib}     % for producing References section
\usepackage[T1]{fontenc}% for producing Polish vowels etc...

\newcommand{\hec}{{\tt HEC22} }
\newcommand{\hece}{{\tt HEC22}}

\newcommand{\vypare}{{\tt VYPAR}}

\newcommand{\ubv}{\hbox{$U\!B{}V$}}
\newcommand{\bv}{\hbox{$B\!-\!V$}}
\newcommand{\ub}{\hbox{$U\!-\!B$}}

\newcommand{\p}{$\pm$}

\newcommand{\m}{$^{\rm m}\!\!.$}

\begin{document}
\countdef\pageno=0
\pageno=3
\def\tit{Forty years of \ubv\ photometry at Hvar}
\def\aut{P. Harmanec \& H. Bo\v{z}i\'c}
\def\str{3--8}

\title{Forty years of \ubv\ photometry at Hvar}

\author{Petr HARMANEC$^1$ and Hrvoje BO\v{Z}I\'C$^2$
%\email{hbozic@geof.hr; hec@sirrah.troja.mff.cuni.cz}
\vspace{2mm}\\
\it $^1$Astronomical Institute of the Charles University,
    Faculty of Mathematics and Physics,\\
\it V~Hole\v{s}ovi\v{c}k\'ach~2, CZ-180~00~Praha~8, Czech Republic\\
\it $^2$Hvar Observatory, Faculty of Geodesy, University of Zagreb,\\
\it Ka\v{c}i\'{c}eva 26, HR--10000 Zagreb, Croatia\\
}

\maketitle

\begin{abstract}
The history of the program of systematic \ubv\ photometric monitoring of
Be stars, binaries, CP stars and some other targets is briefly
summarized. It is shown that a careful data homogenization, reduction
and transformation to the standard Johnson system can be carried out
successfully even at a station nearly at the sea level when some strict
measures are taken.
\end{abstract}

\keywords{\ubv\ photometry - Be stars - binaries}

\section{Motivation}
\begin{figure}[ht]
\begin{minipage}[b]{0.45\linewidth}
\centering
\includegraphics[width=\textwidth, angle=270]{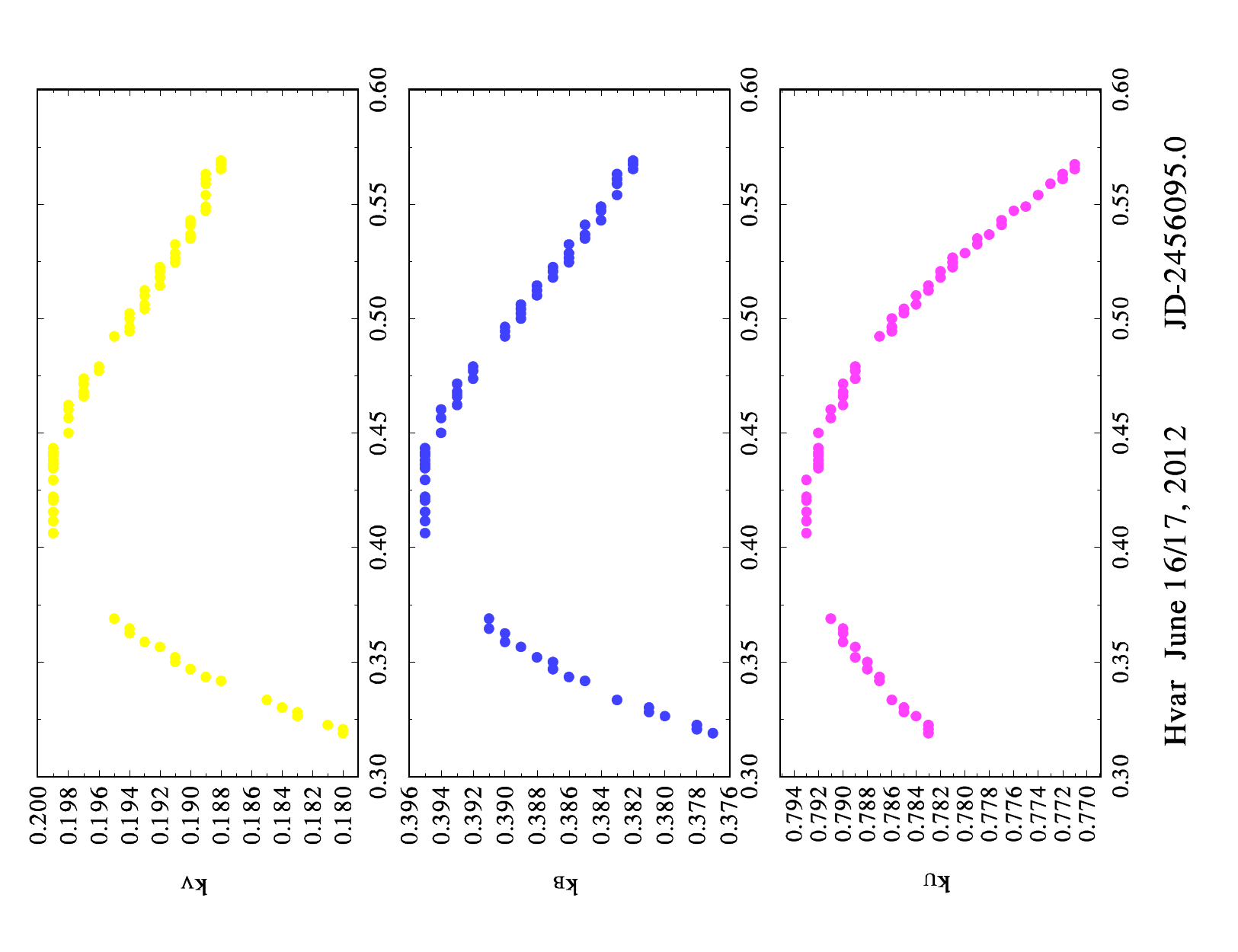}
\caption{The variation of extinction coefficients during one
observing night at Hvar. We found similar changes even for observations
from Sutherland or La~Silla.
}\label{ext}
\end{minipage}
\hspace{0.5cm}
\begin{minipage}[b]{0.45\linewidth}
\centering
\includegraphics[width=\textwidth]{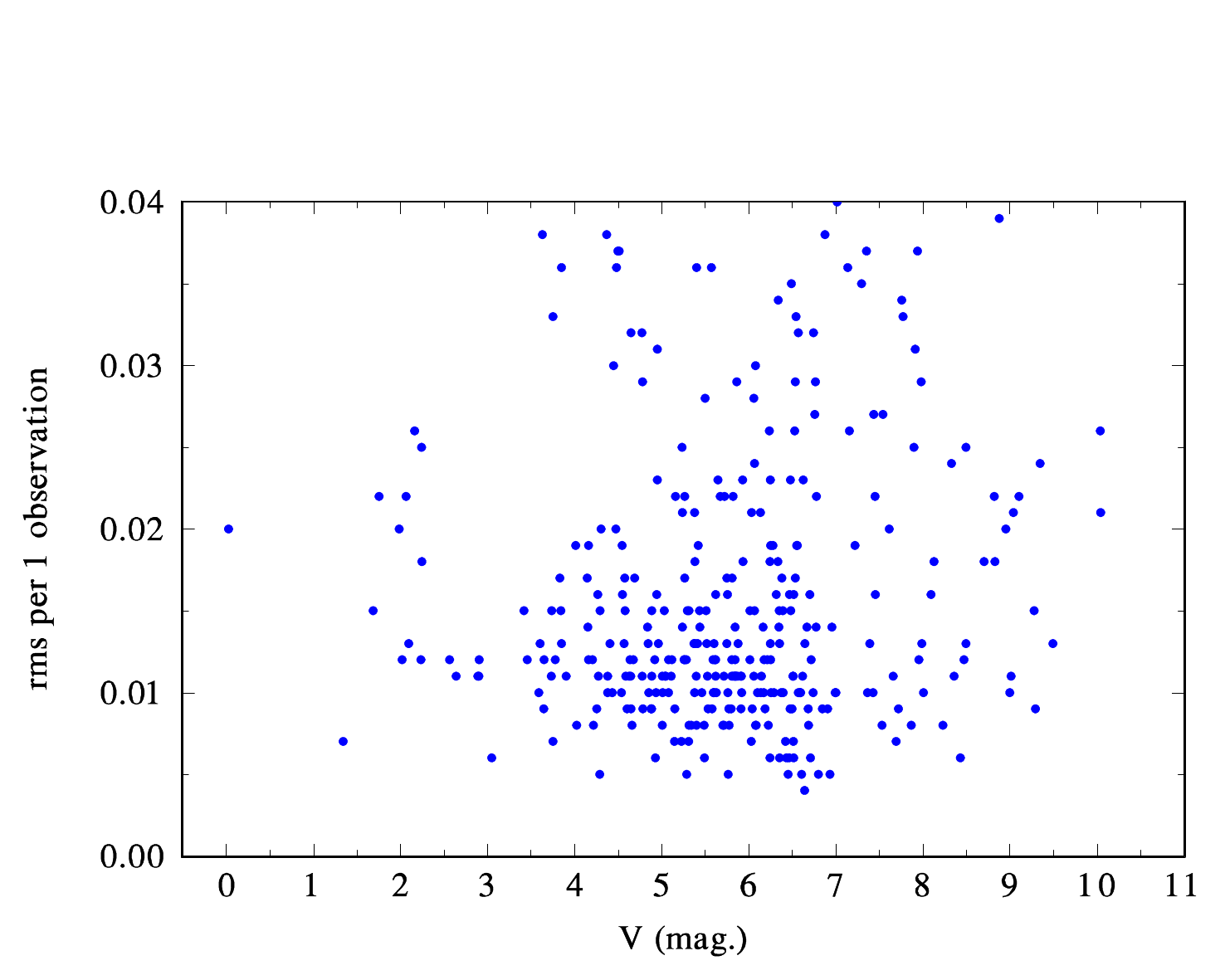}
\caption{A plot of the rms error per 1 observation for all observations
from the Hvar differential archive from the years 1972 to 2012.
The lower envelope of this plot shows that non-variable stars have
the rms errors per 1 observation less than 0\m01 for all stars brighter
than, say 8$^{\rm th}$ magnitude. For data obtained after 1990, this limit is
probably a bit lower.
}\label{rmsv}

\end{minipage}
\end{figure}

Our original goal was to discover some eclipsing binaries among Be stars
to support of the binary hypothesis of the origin of the Be phenomenon
(Harmanec et al. 1972; K\v{r}\'\i\v{z} and Harmanec 1975).
The \ubv\ observing program was started immediately after the installation
of the 0.65-m reflector on July 29, 1972 and has continued until now,
with a number of Croatian, Czech and other observers participating. The
program was parallel to spectroscopic monitoring of Be stars in Ond\v{r}ejov.
Soon we found that the Be stars vary on several timescales, the dominant being
long-term changes on a timescale of years but often with quite a small
amplitude. Therefore, there was a need for a very careful transformation to
the standard Johnson system. The following {\sl non-linear seasonal
transformations} are a must
(see Harmanec, Horn and Juza 1994 and references therein for details):
\begin{eqnarray}
M_{\rm stand.}-M_{\rm instr.}&=&H_1(\bv)+H_2(\ub)\nonumber\\
                             &+&H_3(\bv)^2+H_4(\bv)^3+H_5,
\end{eqnarray}
\noindent where $M$ stands for $V$, $B$, and $U$ and $H_{\rm j}$ are
the seasonal transformation coefficients (different for each passband).
The second crucial step was to homogenize but not to re-define original
Johnson \ubv\ magnitudes of many stars. This way, our comparison, check and
red standard stars can be used as seasonal transformation standards.
The whole package of reduction programs with \hece, \vypare, auxiliary
programs, a detailed manual, and practical examples is available at
{\sl http://astro.troja.mff.cuni.cz:ftp/hec/PHOT}\,. The reduction program
\hec is also able to model the changes of extinction during observing nights,
which greatly improves the accuracy of all-sky photometry, therefore also
the seasonal transformations - see Figure~\ref{ext}.
The second order extinction coefficients are derived among the {\sl seasonal}
transformation coefficients, not every night, since they are given by
the transmission properties of the instrument, not the atmosphere of the Earth.

\section{Statistics}
It is obvious that choosing a site for a photometric observatory,
located only 240~m above the sea level on a relatively small island and
not very far from a city of Hvar is, well, somewhat unusual. We hope we
may, after 40 years of a successful operation, quote the words of the former
director of the Astronomical Institute of the Academy of Sciences in
Ond\v{r}ejov Dr.~L.~Perek. When he first climbed the `goat trail' to the
observatory and looked around, he said: {\sl ``It is a nonsense
to built an observatory here... but what a marvelous one!"}  Our task was
to give sense to this nonsense and
we hope we did... The number of publications in refereed journals and a simple
statistics show that quite clearly.
Our approach led to a `rehabilitation' of the \ubv\ system: Applying
reductions with \hece, we were able to combine the data from Hvar (240~m
above the sea level) and San Pedro M\'artir (2800~m above the sea level)
{\sl without need for any artificial zero-point shifts.}
How well the \ubv\ magnitudes from
different stations can be reproduced is illustrated by Table~1, adopted
from Bo\v{z}i\'c et al. (2007).

\bigskip
%\begin{table}
{\bf Table 1} \ {\scriptsize Mean differential \ubv\ values of the check star
$\varphi$~Her (HD 145389) relative to $\upsilon$~Her from individual
stations and observing seasons, illustrating the level of homogeneity of our
transformations to the standard system.
\begin{center}
\begin{tabular}{rccrcccl}
\hline\noalign{\smallskip}
Station& Mean  epoch   & Year &\ N of & $V$ & \bv & \ub\\
      &(HJD-2400000)&  &          obs. &(mag.)&(mag.)&(mag.)          \\
\noalign{\smallskip}\hline\hline\noalign{\smallskip}
Hvar & 45072.6 &1982&  99\ \ & 4.252\p0.008&\ $-$0.064\p0.008&\ $-$0.248\p0.011\\
Hvar & 51377.2 &1999&   5\ \ & 4.257\p0.007&\ $-$0.062\p0.009&\ $-$0.248\p0.013\\
Hvar & 52305.2 &2002& 139\ \ & 4.254\p0.006&\ $-$0.062\p0.008&\ $-$0.252\p0.008\\
Hvar & 52691.4 &2003&  42\ \ & 4.255\p0.007&\ $-$0.061\p0.010&\ $-$0.255\p0.010\\
Hvar & 53039.0 &2004&  96\ \ & 4.254\p0.005&\ $-$0.061\p0.005&\ $-$0.253\p0.006\\
Hvar & 53370.2 &2005&  78\ \ & 4.257\p0.006&\ $-$0.064\p0.006&\ $-$0.252\p0.006\\
\noalign{\smallskip}\hline\noalign{\smallskip}
San Pedro& 52608.7 &2002&  48\ \ & 4.259\p0.007&\ $-$0.062\p0.007&\ $-$0.247\p0.007\\
San Pedro& 52750.7 &2003&  28\ \ & 4.257\p0.006&\ $-$0.061\p0.007&\ $-$0.249\p0.005\\
\noalign{\smallskip}\hline\noalign{\smallskip}
Hipparcos & 48078.1 &1990&  43\ \ & 4.248\p0.004&all-sky\\
Hipparcos & 48377.8 &1991&  45\ \ & 4.249\p0.005&all-sky\\
Hipparcos & 48700.0 &1992&  24\ \ & 4.249\p0.004&all-sky\\
Hipparcos & 49046.7 &1993&   6\ \ & 4.246\p0.007&all-sky\\
\noalign{\smallskip}\hline\noalign{\smallskip}
Tubitak & 52764.7 &2003&  35 &\ \ 4.253\p0.007&\ $-$0.062\p0.006&\ $-$0.244\p0.009\\
\noalign{\smallskip}\hline\noalign{\smallskip}
\end{tabular}
\end{center}
%\end{table}
}

\begin{figure}
\begin{flushleft}
\includegraphics[scale=1.0,angle=-90]{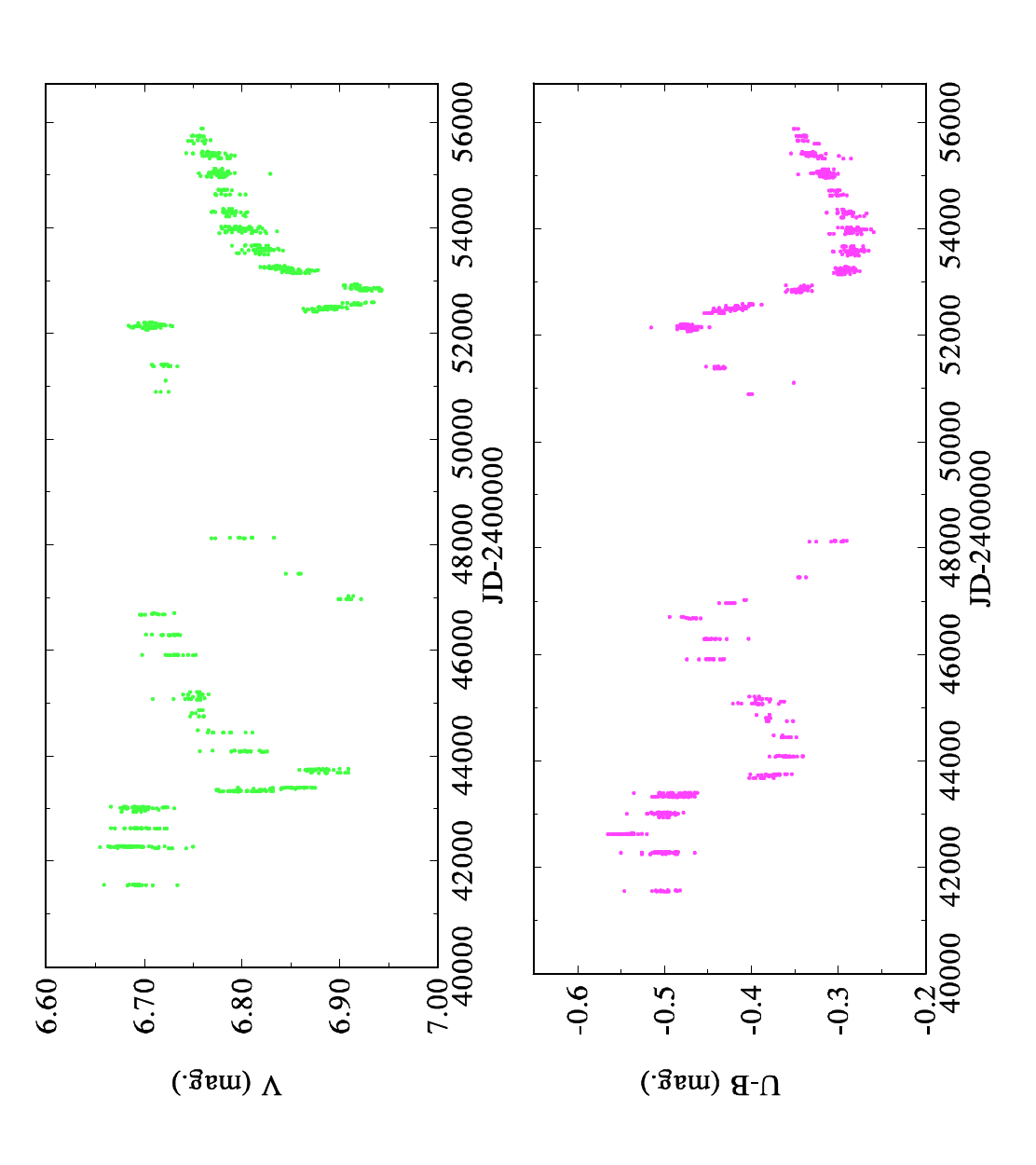}
\caption{Forty years of monitoring of the secular light and colour changes
of the spectroscopic binary with a B7e primary V744~Her = 88~Her,
the very first SB discovered with the Ond\v{r}ejov 2-m telescope.
The data cover three consecutive shell and emission-line episodes and
the plot is based on unedited individual observations from the Hvar archive.
The gap in coverage is due to war years in the former Yugoslavia.}
\label{v744her}
\end{flushleft}
\end{figure}

It turned out that in the years when it was possible to carry
out observations in the course of the whole year (1981 and 1982), thanks
mainly to the enthusiasm of (at that time) youngsters Kre\v{s}o Pavlovski
and Hrvoje Bo\v{z}i\'c, it was possible to obtain usable photometry on more
than 100 nights per year. The current number of about 70 nights per year
is mainly due to the lack of manpower. The latest edition of
the Hvar archive of differential observations from April 2012
contains 75667 observations for 682 different stars. 
Figure~\ref{rmsv} is a plot of a rms error of 1 observation from that
archive. For non-variable stars this error amounts
to less than 0\m01 in $V$, being similar in $B$ and slightly higher in $U$.
\section{Results}
The original instrumentation and reduction strategy are described in the paper
by Harmanec et al. (1977). Results of the first 25 years of observations in the
Be program were summarized in detail by Pavlovski et al. (1997). For the more
recent results in the Be program, see the summary paper by Bo\v{z}i\'c,
Koubsk\'y and Harmanec here.

\begin{figure}
\begin{flushleft}
\includegraphics[scale=0.70,height=58mm]{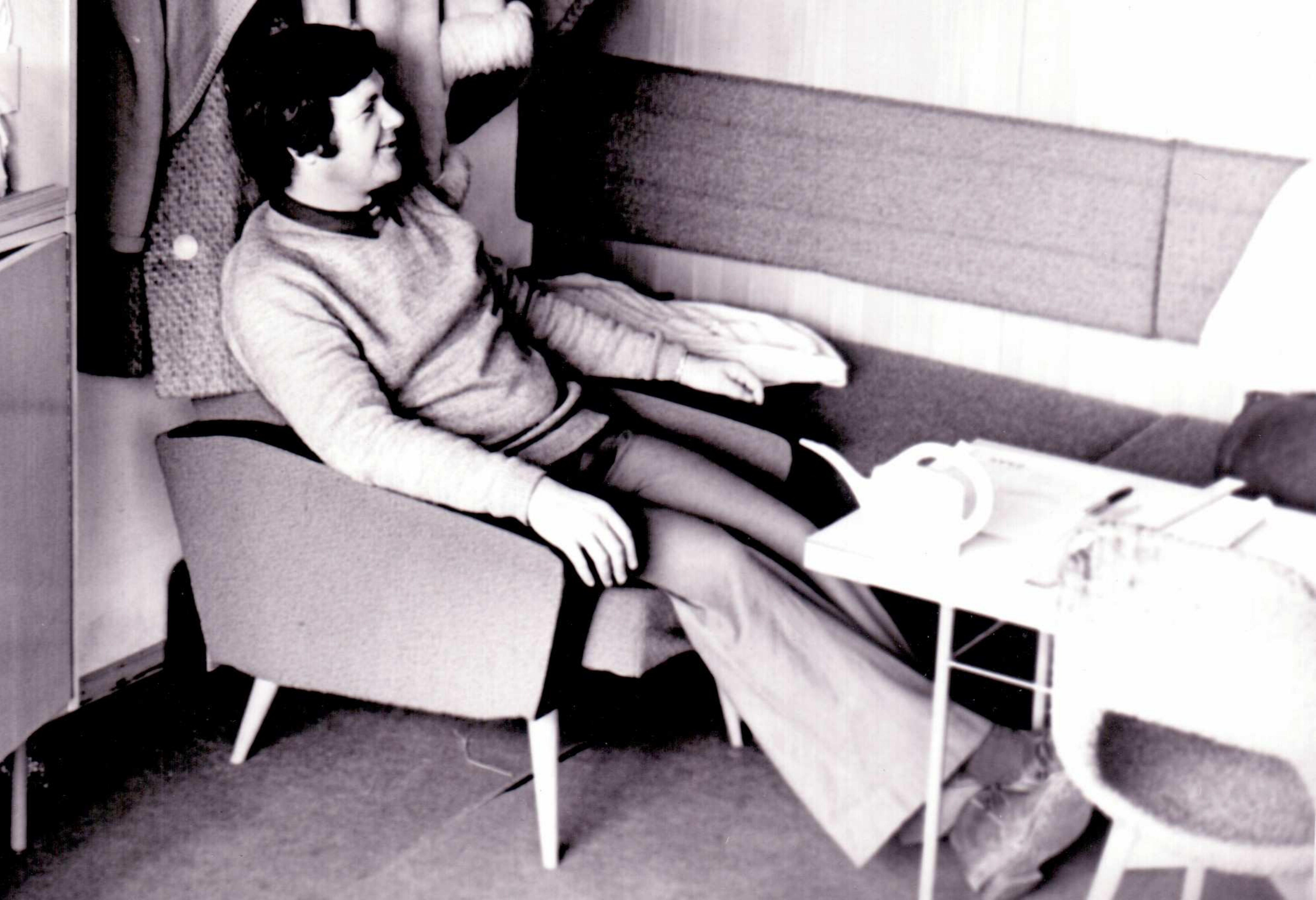}
\includegraphics[scale=0.70,height=58mm]{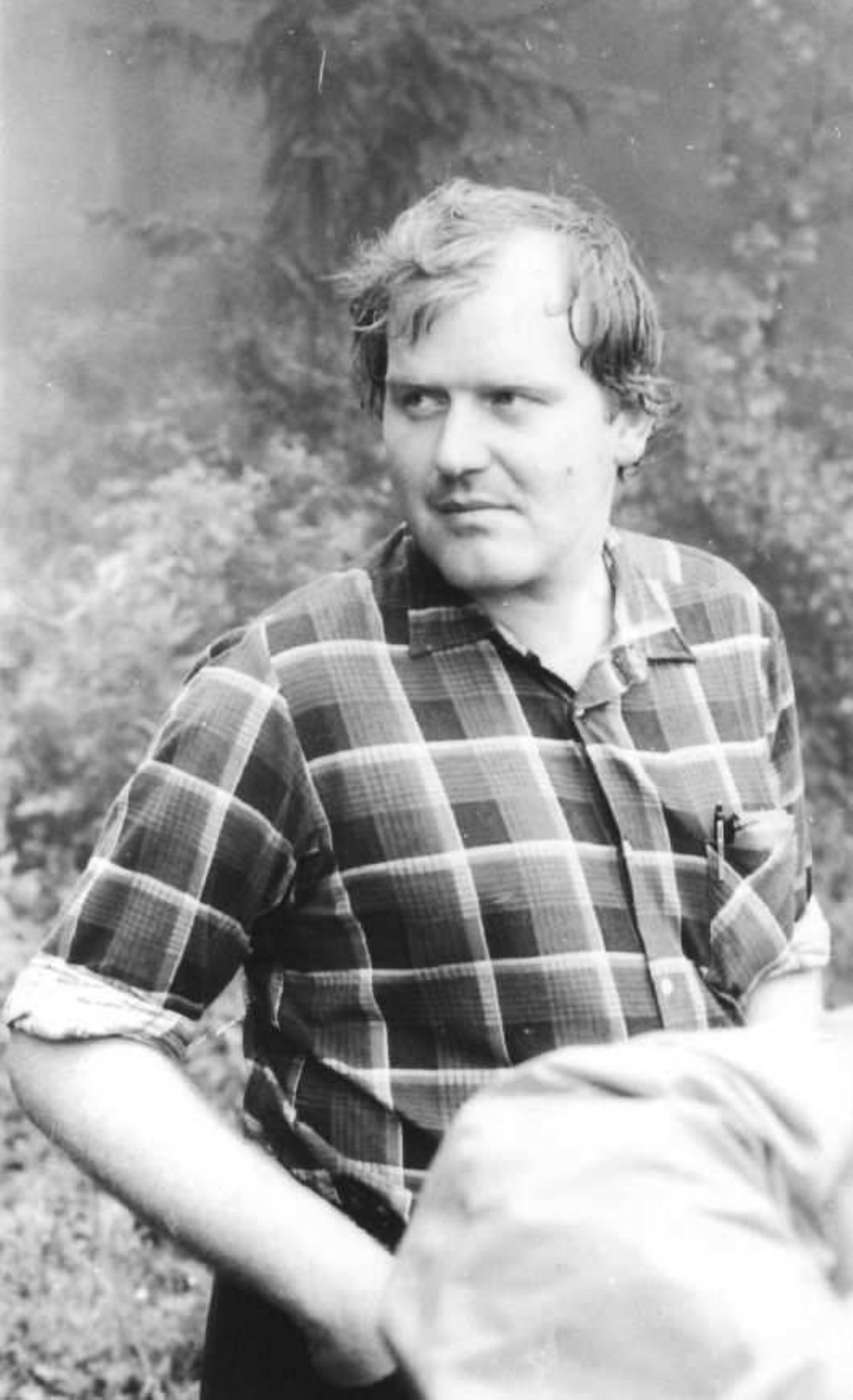}
\caption{To the memory of our friends and colleagues Dr. Ji\v{r}\'\i\ Horn
(left) and Karel Juza (right), who both passed away in 1994. Besides observing
at Hvar, Ji\v{r}\'\i\ wrote the control program of the photometer {\tt PEHVAR}
and the program {\tt VYPAR}, which handles the data archives, while Karel did
the substantial parts of the homogenization of \ubv\ magnitudes after
the first two decades of \ubv\ observations at Hvar.}
\label{kluci}
\end{flushleft}
\end{figure}

The power of systematic observations with a simple instrumentation but
a very careful observational strategy and data reduction is well
illustrated by Fig.~\ref{v744her}. Not many similarly systematic series
of observations exist.

Besides Be stars, some observations were also devoted to observations of CP
stars and in recent years, the main focus has shifted to studies of
astrophysically interesting binary and multiple stars (cf., e.g., Nemravov\'a
et al. here).

We wish to dedicate this contribution to the memory of our good friends and
colleagues Drs.~Ji\v{r}\'\i~Horn and Karel~Juza -- see Fig.~\ref{kluci}.

\section*{Acknowledgements}
The accumulation of the large number of observations would be impossible
without patient and tedious work of a large number of observers, most of them
from Croatia and Czech Republic but also of craftsmen from the Ond\v{r}ejov
mechanical workskop led by K.~Havl\'\i\v{c}ek, Dr. P. Mayer, who designed
the telescope and photometer and a number of other technicians and
servicemen from both countries who could not be mentioned explicitly
due to limited space.
We, however, wish to thank to Dr. V. Ru\v{z}djak, the Director of the Hvar
Observatory, for his continuing support to this project.
The research of PH was supported by the grant P209/10/0715 of the Czech
Science Foundation.

\end{document}